\documentclass[11pt]{article}
\usepackage{amsmath}
\begin{document}
\begin{titlepage}
\vspace*{.5in}

 \begin{center}
{\Large\bf
How to Hide a Cosmological Constant}\\  
\vspace{.4in}
{S{\sc teven} C{\sc arlip}\footnote{\it email: carlip@physics.ucdavis.edu}\\
       {\small\it Department of Physics}\\
       {\small\it University of California}\\
       {\small\it Davis, CA 95616}\\{\small\it USA}}
\end{center}

\vspace{.5in}
\begin{center}
{\large\bf Abstract}
\end{center}
\begin{center}
\begin{minipage}{4.7in}
{\small
Naive calculations in quantum field theory suggest that vacuum fluctuations should induce
an enormous cosmological constant.  What if these estimates are right?  I argue that 
even a huge cosmological constant might be hidden in Planck scale fluctuations of geometry 
and topology---what Wheeler called ``spacetime foam''---while remaining virtually invisible 
macroscopically.}
\end{minipage}
\end{center}
\vspace*{2ex}
\emph{\small Essay written for the Gravity Research Foundation 2019 Awards for Essays on Gravitation\\[.5ex]
Submitted March 22, 2019\\[.5ex]
Winner, Fourth Award}
\end{titlepage}
\addtocounter{footnote}{-1}

Everything gravitates.  The universality of gravity, an expression of the principle of equivalence, 
is a cornerstone of general relativity, and is one of the most precisely tested experimental
phenomena in physics \cite{POE,Touboul}.

But ``everything'' presumably includes vacuum energy, the energy of quantum fluctuations
of empty space, whose gravitational effect should manifest itself as a cosmological constant 
$\Lambda$.  We don't know how to calculate this constant, but simple estimates give answers 
that are between $55$ and $122$ orders of magnitude larger than we observe \cite{Martin}. 
This crisis has been called ``the worst theoretical prediction in the history of physics''  \cite{Hobson}, 
and has led to what can only be called desperate measures: appeals to the anthropic principle 
\cite{Weinberg}, ad hoc nonlocal modifications of general relativity \cite{Kaloper}, and the like.

Here I suggest an alternative.  Perhaps the Universe really \emph{does} have an enormous
cosmological constant \cite{Carlip}.  If we assume homogeneity, this is immediately ruled out by
observation, of course.  But if $\Lambda$ comes from quantum fluctuations  at the Planck
scale, there is no reason to expect homogeneity at that scale.  Could it be that the effects
of vacuum energy are absorbed by large curvature and complex topology at tiny scales, while 
averaging to near zero macroscopically?

This is not a new idea: more than 60 years ago, Wheeler proposed a similar notion of ``spacetime
foam'' at the Planck scale \cite{Wheeler}.  But today we have better tools for investigating this 
possibility.  While we do not yet have an answer, I will argue below that the picture is at least 
plausible.

\section{Initial data and evolution}

It's convenient to start with the initial value formalism of general relativity.  Choose a time slice
 $\Sigma$, a cross-section of spacetime at a ``moment of time.''  The initial data are then
a spatial metric $g_{ij}$ on $\Sigma$ and an extrinsic curvature $K^i{}_j$, essentially the time 
derivative of the metric.  The trace $K=K^i{}_i$, called the ``expansion,'' is the local Hubble 
constant, the fractional rate of change of the local volume element with time.  

These data are not independent.  They must satisfy two constraints,
\begin{align}
&R + K^2 - K^i{}_jK^j{}_i - 2\Lambda = 0 \, , \nonumber \\
&D_i(K^i{}_j - \delta^i_jK) = 0 \, ,\label{a1}
\end{align}
where $R$ is the scalar curvature of the metric $g_{ij}$ and $D_i$ is the compatible covariant 
derivative.  Any solution of these constraints can be evolved to form at a full four-dimensional 
spacetime.

Now suppose this initial slice has a complicated geometric and topological structure at the 
Planck scale.  Any topological three-manifold $\Sigma$ has a ``prime decomposition'' 
\begin{align}
\Sigma = \Sigma_1\#\Sigma_2\#\dots\#\Sigma_N \, ,
\label{a2}
\end{align}
where the ``prime factors'' $\Sigma_\alpha$ are elementary pieces of topology, joined by 
cutting out solid balls and identifying the resulting edges \cite{top}.  It has recently been 
shown that this ``connected sum''  decomposition extends to initial data as well: generic initial 
data on the prime factors can also be consistently glued together \cite{Chrusciel,Chruscielb}. 

We now make a crucial observation.  The constraints (\ref{a1}) are time-reversal invariant: if
$(g_{ij},K^i{}_j)$ is a solution on a prime factor $\Sigma_\alpha$, so is $(g_{ij},-K^i{}_j)$.  So if we 
randomly combine enough Planck-size pieces of topology---using any criteria we like, as long as
we don't assume a microscopic arrow of time---positive and negative extrinsic curvature should 
occur equally often.   A cubic centimeter contains $10^{100}$ such Planck-size volumes;
with any sensible averaging procedure, the average $\langle K^i{}_j\rangle$ will approach zero.  
We thus obtain spacetimes in which the cosmological constant has a huge effect at the Planck 
scale---each elementary piece will typically have an enormous expansion---but becomes hidden 
macroscopically.  

This isn't yet quite good enough.  To match our observed Universe, we also want the average intrinsic
curvature $R$ to be small.  This will be true for an infinite class of initial data, but it's not currently
clear why this feature should be preferred.  There is some evidence, though, that as in
ordinary closed FLRW cosmology, a universe that starts with high average intrinsic curvature 
will evolve toward small curvature \cite{Carlip}.

So far, I've focused on initial data.  The crucial question is whether time evolution preserves 
these nice features.  A full answer probably requires a much better grasp of quantum gravity, 
and in particular a better description of Planck-scale fluctuations.  Naively, one might make 
two rather different guesses:

 -- Expanding regions grow in time while contracting regions shrink, so over time the 
expanding regions should dominate.
 
 -- But nothing in this construction picks out a ``preferred'' initial time.  If the foamy
structure arises from quantum fluctuations, it should replicate itself: expanding regions 
should fill up with new curvature fluctuations.

We don't yet know which, if either, of these intuitions is correct.  Classically, there is  
evidence that if we start with a complex initial data structure (\ref{a2}), the resulting
spacetime near the initial surface can be foliated by time slices that all have vanishing average
expansion $\langle K\rangle$ \cite{Carlip}.  This classical evolution can't continue indefinitely, though; 
initial data constructed this way typically has regions that collapse into  black-hole-like singularities 
\cite{Burkhart}.  Here, again, we must appeal to Wheeler's insistence that a full quantum theory 
of gravity should eliminate such singularities.

\section{Quantum theory}

My argument so far has been semiclassical, using quantum theory to generate Planck scale
structure but otherwise viewing evolution classically.  We can say much less about the full quantum
treatment, although it would be interesting to try to connect this picture to Hawking's Euclidean
spacetime foam \cite{Hawking} and to AdS/CFT discussions of sums over topologies \cite{Farey}.  

There is, however, one feature of quantum gravity that offers some hope.   The formalism
I've used contains two ingredients: constraints, which we understand well, and 
evolution equations, which we don't.  In many approaches to quantum gravity, though, the 
constraints are everything.  The Wheeler-DeWitt equation, for instance, is just the constraint 
(\ref{a1}), written as an operator equation \cite{WdW}.   This may merely hide difficult questions 
about time evolution, part of the notorious ``problem of time'' in quantum gravity \cite{Kuchar}.  
But it suggests a potentially fruitful rephrasing of the question: does a typical solution of the 
Wheeler-DeWitt equation have a ``foamy'' structure on an arbitrary time slice?
 
One might also ask how to recover macroscopic general relativity from this microscopic picture.
This is another famously hard problem, that of averaging local inhomogeneities \cite{Clarkson}.  
Here, though, methods of effective field theory may help \cite{EFT}.  As long as the microscopic 
structure preserves diffeomorphism invariance,  standard arguments tell us  that the effective 
macroscopic action will still be that of general relativity, although with renormalized couplings.

I claimed at the beginning of this essay that existing approaches to the cosmological constant are
desperate measures.  I will leave it to the reader to decide whether this is another.  I would argue, 
though, that this proposal has one unusually nice feature.  We \emph{know} that Planck scale 
quantum fluctuations are essential: they generate the vacuum energy that causes the problem
in the first place.  If these ideas are right, the same fluctuations also provide the cure, with no 
extra ingredients needed.  In 1957, Wheeler wrote,
\begin{quote}
\dots it is essential to allow for fluctuations in the metric and gravitational interactions in any
proper treatment of the compensation problem---the problem of compensation of ``infinite'' energies that
is so central to the physics of fields and particles \cite{Wheelerb}.
\end{quote}
Perhaps he was right.

\end{document}